\documentclass[%
reprint,
superscriptaddress,
showpacs,
preprintnumbers,
amsmath,amssymb,
prc,
floatfix, ]%
{revtex4-1}

\usepackage{color}

\usepackage{graphicx}% Include figure files
\usepackage{dcolumn}% Align table columns on decimal point
\usepackage{bm}% bold math
\usepackage[dvipdfmx,bookmarks=true,colorlinks,%
            citecolor=blue,linkcolor=blue,anchorcolor=blue,filecolor=blue,urlcolor=blue,%
           ]{hyperref}
\begin{document}

%\begin{CJK*}{GBK}{}

%\begin{CJK*}{UTF8}{gbsn}

\title{Systematic study of breakup effects on complete fusion 
       at energies above the Coulomb barrier}

\author{Bing Wang}% (王兵)}%
%\email{121wb163.com}
 \affiliation{Department of Physics, Zhengzhou University, Zhengzhou 
              450001, China}
 \author{Wei-Juan Zhao}% (赵维娟)}%
%\email{zwj@zzu.edu.cn}
 \affiliation{Department of Physics, Zhengzhou University, Zhengzhou 
              450001, China}
 \author{P. R. S. Gomes}%
%\email{paulogom@if.uff.br }
 \affiliation{Instituto de F\'isica, Universidade Federal Fluminense, 
              Avenida Litoranea s/n, Gragoat\'a, Niter\'oi, Rio de Janeiro, 
              24210-340, Brazil}
\author{En-Guang Zhao}% (赵恩广)}%
%\email{egzhao@mail.itp.ac.cn}
 \affiliation{State Key Laboratory of Theoretical Physics,
              Institute of Theoretical Physics, Chinese Academy of Sciences, 
              Beijing 100190, China}
 \affiliation{Center of Theoretical Nuclear Physics, National Laboratory
              of Heavy Ion Accelerator, Lanzhou 730000, China}
\author{Shan-Gui Zhou}% (周善贵)}%
 \email{sgzhou@itp.ac.cn}
%\homepage{http://www.itp.ac.cn/~sgzhou}
 \affiliation{State Key Laboratory of Theoretical Physics,
              Institute of Theoretical Physics, Chinese Academy of Sciences, 
              Beijing 100190, China}
 \affiliation{Center of Theoretical Nuclear Physics, National Laboratory
              of Heavy Ion Accelerator, Lanzhou 730000, China}
 \affiliation{Center for Nuclear Matter Science, Central China Normal University, 
              Wuhan 430079, China}

 \date{\today}

\begin{abstract}
A large number of complete fusion excitation functions of reactions including 
the breakup channel were measured in recent decades, especially in the last few 
years. It allows us to investigate the systematic behavior of the breakup 
effects on the complete fusion cross sections. To this end, we 
perform a systematic study of the breakup effects on the complete fusion 
cross sections at energies above the Coulomb barrier. 
The reduced fusion functions $F(x)$ are compared with the universal 
fusion functions which are used as a uniform standard reference. 
The complete fusion cross sections at energies above the Coulomb barrier
are suppressed by the breakup of projectiles. 
This suppression effect for reactions induced by the same projectile 
is independent of the target and 
mainly determined by the lowest energy breakup channel of the projectile.
There holds a good exponential relation between the suppression factor
and the energy corresponding to the lowest breakup threshold.
\end{abstract}

%25.60.Pj 	Fusion reactions
%24.10.-i 	Nuclear reaction models and methods
%25.70.Mn 	Projectile and target fragmentation
%25.70.Jj 	Fusion and fusion-fission reactions

%25.60.Gc 	Breakup and momentum distributions
 
\maketitle

%\end{CJK*}
\section{Introduction}
\label{sec:introduction}

In recent years, the investigation of breakup effects on fusion reactions in 
heavy-ion collisions around the Coulomb barrier has been a subject of intense 
experimental and theoretical interests
\cite{Canto2006_PR424-1,Keeley2007_PPNP59-579,Back2014_RMP86-317}. Various 
processes can take place after the projectile breaks up. One is the incomplete 
fusion (ICF) in which part of the fragments is absorbed by the target. 
When all the fragments fuse with the target, the process is called 
sequential complete fusion (SCF). From the experimental point of view, the SCF 
cannot be distinguished from the direct complete fusion (DCF) in which the 
whole projectile fuses with the target without breakup. Therefore, only the 
complete fusion (CF) cross section, which includes both DCF and SCF 
cross sections, can be measured. 

The total fusion (TF) cross section is the sum 
of the CF and ICF cross sections, $\sigma_{\rm TF} \equiv \sigma_{\rm 
CF} + \sigma_{\rm ICF}$. Experimentally, it is difficult to measure separately 
ICF and CF cross sections. Especially for light reaction systems, the excited 
compound nucleus emits charged particles during the cooling process. The 
residues from ICF cannot be distinguished from those from CF, and hence only 
the TF cross section can be measured. The situation is different in the case of 
heavy reaction systems, because the decay of the excited compound nucleus 
through the emission of charged particles can be negligible and the 
separate measurements of the CF cross section can be achieved. Many 
measurements of the CF cross sections have been performed 
\cite{Broda1975_NPA248-356,Tripathi2002_PRL88-172701,
Liu2005_EPJA26-73,Gomes2009_NPA828-233,
Shrivastava2009_PRL103-232702,Fang2013_PRC87-024604,Dasgupta2004_PRC70-024606,
Rath2009_PRC79-051601R,Tripathi2005_PRC72-017601, 
Gasques2009_PRC79-034605,Kalita2011_JPG38-095104}.

Several methods have been adopted to investigate the influence of the breakup 
on 
fusion reactions around the Coulomb barrier 
\cite{Hagino2000_PRC61-037602,Diaz-Torres2002_PRC65-024606,
Gomes2011_PRC84-014615,Gomes2012_JPG39-115103,Sargsyan2012_PRC86-054610,
Boselli2014_JPG41-094001}. One 
of the most widely employed approaches is to compare the data with either the 
predictions 
of coupled channel (CC) calculations without the breakup channels  
\cite{Dasgupta2004_PRC70-024606,Rath2009_PRC79-051601R,Kumawat2012_PRC86-024607,
Rath2012_NPA874-14,Pradhan2011_PRC83-064606,Palshetkar2014_PRC89-024607,
Rath2013_PRC88-044617} or the predictions of a single barrier penetration 
model (SBPM) 
\cite{Tripathi2002_PRL88-172701,Tripathi2005_PRC72-017601, 
Gasques2009_PRC79-034605,Kalita2011_JPG38-095104}. It was found that the CF 
cross sections are suppressed at energies above the Coulomb barrier. 
In Refs.~\cite{Gasques2009_PRC79-034605,Dasgupta2010_NPA834-147c}, it was 
concluded that the CF suppression for the reactions involving ${}^{6}$Li, 
${}^{7}$Li, and ${}^{10}$B projectiles is almost independent of the target 
charge by comparing the CF data with the predictions of CC or SBPM calculations,
and the suppression shows a remarkably consistent correlation with the 
breakup threshold energy. Sargsyan {\it et al.} 
\cite{Sargsyan2012_PRC86-054610} 
investigated the systematic behavior for the CF suppression as a function of 
the target charge and bombarding energy by using the quantum diffusion 
approach. In Ref. \cite{Gomes2011_PRC84-014615}, the influence of breakup 
effects on CF cross sections for ${}^{9}$Be induced reactions were discussed by 
applying the universal fusion function formalism and it was concluded that 
there is not a clear systematic behavior of the CF suppression as a function of 
the target charge.

As mentioned above, the conclusions concerning the CF 
suppression obtained by different methods are different. Therefore, further 
systematic study of the influence of the breakup on CF cross 
sections is needed. Particularly, in the last few years, new 
experiments with weakly bound nuclei were performed and 
the corresponding CF cross sections were measured
\cite{Pradhan2011_PRC83-064606,Rath2012_NPA874-14,Kumawat2012_PRC86-024607, 
Rath2013_PRC88-044617,Palshetkar2014_PRC89-024607}. Moreover, evidences 
of breakup for tightly bound projectiles (${}^{11}$B, ${}^{12,13}$C and 
${}^{16}$O) were 
also found \cite{Singh2008_PRC77-014607,Palshetkar2010_PRC82-044608, 
Kalita2011_JPG38-095104,Yadav2012_PRC86-014603,Yadav2012_PRC85-034614}. 
It allows us to explore the breakup effects on CF cross sections in 
a wider range.  

In order to perform a systematic study of the breakup effects on CF cross 
sections, it is necessary to reduce the data to eliminate the 
geometrical factors in different reaction systems 
\cite{Gomes2010_NPA834-151c}. 
After the reduction, the data should be compared with the theoretical 
predictions without taking into account the coupling of the breakup channel.

Several reduction methods have been used to reduce the data
\cite{Beckerman1982_PRC25-837,DiGregorio1989_PRC39-516,Gomes2005_PRC71-017601,
Canto2009_JPG36-015109, Canto2009_NPA821-51,Wolski2013_PRC88-041603R}.
In this paper, the one proposed in 
Refs.~\cite{Canto2009_JPG36-015109,Canto2009_NPA821-51}, which can eliminate 
completely the geometrical factors and static effects of the potential between 
the two nuclei, is employed to reduce the CF data. We will explore the 
influence of the breakup on CF cross sections at energies above the Coulomb 
barrier, because the coupling channel effects except the breakup do 
not play a significant role on fusion cross sections in this energy range 
\cite{Leigh1995_PRC52-3151,Zhang2011_NPA864-128}.
Meanwhile, in order to avoid using a very large diffuseness parameter 
%, which is not compatible with the one extracted 
%from elastic and inelastic scattering data, 
in the Woods-Saxon potential 
\cite{Newton2004_PRC70-024605,Newton2004_PLB586-219}, 
we choose the double folding and parameter-free S\~ao Paulo potential (SPP) 
\cite{CandidoRibeiro1997_PRL78-3270,Chamon1997_PRL79-5218,
Chamon2002_PRC66-014610} 
as the interaction potential between the projectile and the target. 
The SPP has been widely and successfully used in the study of heavy-ion 
reactions 
\cite{Gasques2004_PRC69-034603,Crema2005_PRC72-034610,Gomes2009_PRC79-027606,
Shorto2009_PLB678-77,Lei2012_PRC86-057603,Yang2013_PRC87-014603}.
The universal fusion function (UFF) 
\cite{Canto2009_JPG36-015109,Canto2009_NPA821-51} will be used as a uniform 
standard reference with which the reduced data can be 
compared directly. 
%The deviations from CF data with respect 
%to the UFF at energies above the Coulomb barrier are mainly owing to the 
%effects of breakup coupling on CF cross sections.

The present paper is organized as follows. In Sec.~\ref{sec:methods} the 
method used to eliminate geometrical factors and static effects of the data and 
the SPP are introduced. 
This method is applied to analyze the data of different projectiles induced 
reactions in Sec.~\ref{sec:results} where the systematics of the suppression
effects from the breakup channel will be investigated.
A summary is given in Sec.~\ref{sec:summary}.

\section{Methods}
\label{sec:methods}

In order to study the systematic behavior of the breakup effects on CF 
cross sections, it is necessary to eliminate completely the geometrical factors 
and static effects of the potential between the two nuclei. We adopt the method 
proposed in Refs. \cite{Canto2009_JPG36-015109,Canto2009_NPA821-51}. According 
to this prescription, the fusion cross section and the collision energy are 
reduced to a dimensionless fusion function $F(x)$ and a dimensionless variable 
$x$,  
\begin{equation}\label{eq:1}
F(x) = \frac{2E_{\rm c.m.}}{R_{\rm B}^2\hbar\omega}\sigma_{\rm F}, \quad
   x = \frac{ E_{\rm c.m.}-V_{\rm B}}{\hbar\omega},
\end{equation}
where $E_{\rm c.m.}$ is the collision energy
in the center of mass frame, $\sigma_{\rm F}$ is the fusion cross section, 
and $R_{\rm B}$, $V_{\rm B}$, and $\hbar\omega$ 
denote the radius, height, and curvature of the barrier which is 
approximated by a parabola. The barrier parameters $R_{\rm
B}$, $V_{\rm B}$, and $\hbar\omega$ are obtained from the SPP.

In the model of SPP, the nuclear interaction $V_{\rm N}$ is given as 
\cite{CandidoRibeiro1997_PRL78-3270,Chamon1997_PRL79-5218,
Chamon2002_PRC66-014610}
\begin{equation} 
   V_{\rm N}(R,E_{\rm c.m.}) \approx V_{\rm 
F}(R)\exp\left(-\frac{4v^2}{c^2}\right),
\end{equation}
where $V_{\rm F}$ is the double-folding potential obtained by using 
the density distributions of the nuclei. The two-parameter Fermi 
distribution is used to describe the densities of the nuclei. $c$ 
denotes the speed of light, and $v$ the relative velocity between the 
projectile 
and the target,
\begin{equation} 
 v(R,E_{\rm c.m.}) =\sqrt{2[E_{\rm c.m.}-V_{\rm C}(R)-V_{\rm N}(R,E_{\rm 
c.m.})]/\mu}.
\end{equation}
$V_{\rm C}$ is the Coulomb potential which is also calculated through a folding 
procedure. $\mu$ is the reduced mass of the reaction system in question. 
 
The reduction method given by Eq.~(\ref{eq:1}) is inspired by the Wong's formula
\cite{Wong1973_PRL31-766}, % for penetrating a single parabolic barrier, 
%Neglecting the variation of the barrier radius and curvature with 
%angular momentum, an analytic expression which is called Wong's formula for 
%fusion cross section is obtained as \cite{Wong1973_PRL31-766}
\begin{equation}\label{eq:2}
  \sigma_{\rm F}^{\rm W}(E_{\rm c.m.}) = \frac{R_{\rm B}^2\hbar\omega}{2E_{\rm
        c.m.}}\ln\left[1+\exp\left(\frac{2\pi(E_{\rm c.m.}-V_{\rm 
        B})}{\hbar\omega}\right)\right].
\end{equation}
If the fusion cross section can be 
accurately described by the Wong's formula, 
the $F(x)$ reduces to
\begin{equation}\label{eq:3}
   F_{0}(x) = \ln\left[1+\exp(2\pi x)\right],
\end{equation}
which is called the UFF \cite{Canto2009_JPG36-015109,Canto2009_NPA821-51}. Note 
that $F_{0}(x)$ is a general function of the 
dimensionless variable $x$ and independent of reaction systems. It is well 
known that the Wong's formula has limitations and does not describe properly 
the behavior of fusion cross sections of light systems at sub-barrier energies. 
However, in the present work we are dealing with energies above the barrier, an 
energy region where the Wong's formula can be applied.
\label{modification:standard}
In particular, when $x>1$, one has $ F_{0}(x) \approx 2 \pi x$. 
Then the fusion cross section is calculated as
\begin{equation}\label{eq:ref}
 \sigma_{\rm F} = \pi R_{\rm B}^2 (E_{\rm c.m.}-V_{\rm B})/E_{\rm c.m.}.
\end{equation}
Thus, fusion becomes independent on the width of the barrier
and can be described as the absorption by a black disc of radius $R_{\rm B}$. 
Partial fusion, producing the reduction of the complete fusion, 
can be visualized as coming from one of the breakup fragments 
not falling into the black disc. \label{modification:Fx} 
So, the $F_{0}(x)$ is used as a uniform 
standard reference to explore the breakup effects on CF cross sections. 

Using the above reduction procedure, the CF data for different reaction systems 
can be compared directly and the systematics can be investigated. 
Deviations of the fusion function, if exist, from the UFF at energies above 
the Coulomb barrier mainly arise from the effects of the breakup on CF cross 
sections 
\cite{Leigh1995_PRC52-3151,Zhang2011_NPA864-128}, because inelastic excitations 
and transfer channel couplings are not important at energies above the Coulomb 
barrier. 
This is also the reason why one does not need to renormalize the experimental 
fusion functions, as prescribed by Canto {\it et al.} 
\cite{Canto2009_JPG36-015109, 
Canto2009_NPA821-51}. 

\section{Results and Discussions}
\label{sec:results}

\begin{figure}
\centering{ \includegraphics[width=0.85\columnwidth]{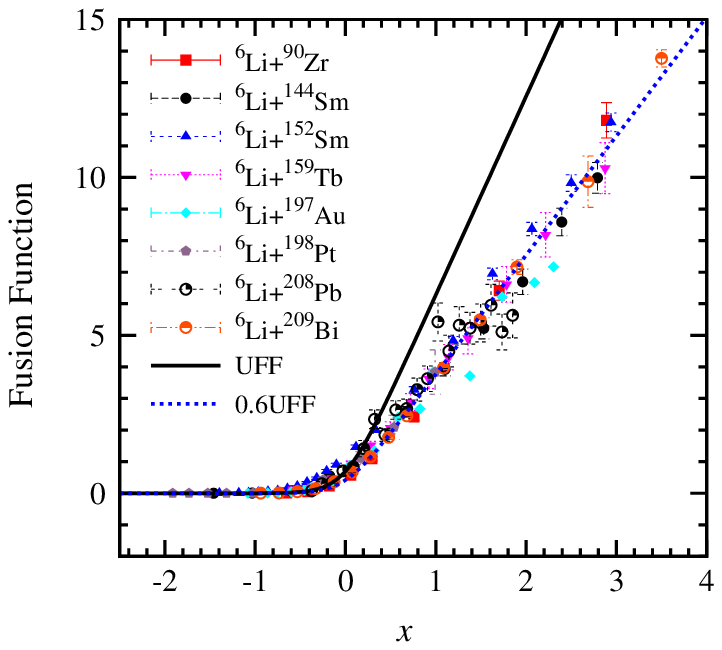}}
\caption{(Color online) The complete fusion function $F(x)$ for the weakly 
bound projectile ${}^{6}$Li on different target nuclei as a function of $x$. 
The solid line represents the UFF [Eq. (\ref{eq:3})] and the dotted
line is the UFF multiplied by the suppression factor $F_\mathrm{B.U.}= 0.60$ 
[cf. Eq. (\ref{eq:4})]. The experimental values are taken from 
Refs.~\cite{Kumawat2012_PRC86-024607}      (${}^{90}$Zr),
\cite{Rath2009_PRC79-051601R}        (${}^{144}$Sm),
\cite{Rath2012_NPA874-14}            (${}^{152}$Sm),
\cite{Pradhan2011_PRC83-064606}      (${}^{159}$Tb), 
\cite{Palshetkar2014_PRC89-024607}   (${}^{197}$Au),
\cite{Shrivastava2009_PRL103-232702} (${}^{198}$Pt),
\cite{Liu2005_EPJA26-73}             (${}^{208}$Pb), and
\cite{Dasgupta2004_PRC70-024606}     (${}^{209}$Bi).}\label{fig:Li6}
\end{figure}

In the last few years, CF cross sections for many reactions involving weakly 
bound nuclei have been measured 
\cite{Pradhan2011_PRC83-064606,Rath2012_NPA874-14,Kumawat2012_PRC86-024607, 
Rath2013_PRC88-044617,Palshetkar2014_PRC89-024607}. The suppression of complete 
fusion 
cross sections above the Coulomb barrier due to the breakup has been observed 
in 
these reaction systems. More interestingly, for 
tightly bound projectiles (${}^{11}$B, ${}^{12,13}$C and ${}^{16}$O), the 
evidence for incomplete fusion has been also found 
\cite{Singh2008_PRC77-014607,Palshetkar2010_PRC82-044608, 
Kalita2011_JPG38-095104,Yadav2012_PRC86-014603,Yadav2012_PRC85-034614}. 
This fact allows us to explore the influence of the breakup on CF cross 
sections 
in a wider range of projectiles and breakup threshold energies. We collected 
the 
CF data for reactions induced by ${}^{6,7}$Li, ${}^{9}$Be, ${}^{10,11}$B, 
${}^{12,13}$C, and ${}^{16}$O which are shown in Table \ref{tab:SF}. Based on 
the reduction method mentioned above, we first investigate the influence of 
breakup effects on the CF cross sections for each projectile. Then we study the 
systematic behavior of the suppression factors for different projectiles.

\begin{table}\label{modification:table}
\caption{ %
The reactions studied in this work. The first and last column denote the 
projectile (Proj.) and target (Targ.) of the reaction, respectively. The second 
column represents the lowest breakup threshold $E_{\rm B.U.}$ (in MeV) for the 
projectile. The suppression factors $F_{\rm B.U.}$ obtained by fitting the CF 
functions are listed in the third column. The $F_{\rm B.U.}^{\rm em}$ 
denotes the suppression factor obtained from the empirical formula 
(\ref{eq:E_BU}). 
The suppression factors taken from the literature are given in the fifth column 
and the corresponding references are shown in the last column.  
}
\begin{ruledtabular}
\begin{tabular}{cccccr}
  Proj.   & $E_{\rm B.U.}$  & $F_{\rm B.U.}$ &  $F_{\rm B.U.}^{\rm em}$    & 
$F_{\rm B.U.}$  & Targ.   \\ 
          & (MeV)           & (fit)      &   [Eq. (\ref{eq:E_BU})]        & 
 (Refs.)         &        \\
 \hline\noalign{\vskip3pt}
 $^{6}$Li     & $1.474$      & $0.60$  & $0.601$ & $0.66\pm 0.08$ & $^{90}$Zr 
\cite{Kumawat2012_PRC86-024607}        \\
              &              &         &         & $0.68$         & $^{144}$Sm 
\cite{Rath2009_PRC79-051601R}          \\
              &              &         &         & $0.72\pm 0.04$ & $^{152}$Sm 
\cite{Rath2012_NPA874-14}              \\
              &              &         &         & $0.66\pm 0.05$ & $^{159}$Tb 
\cite{Pradhan2011_PRC83-064606}         \\
              &              &         &         & $0.65\pm 0.23$ & $^{197}$Au 
\cite{Palshetkar2014_PRC89-024607}      \\
              &              &         &         &                & $^{198}$Pt 
\cite{Shrivastava2009_PRL103-232702}     \\
              &              &         &         &                & $^{208}$Pb 
\cite{Liu2005_EPJA26-73}                 \\
              &              &         &  &$0.66^{+0.05}_{-0.04}$ &$^{209}$Bi 
\cite{Dasgupta2004_PRC70-024606}          \\
 $^{7}$Li     & $2.467$      & $0.67$  & $0.690$ & $0.75\pm 0.04$ & $^{144}$Sm 
\cite{Rath2013_PRC88-044617}              \\
              &              &         &         & $0.75\pm 0.04$ & $^{152}$Sm 
\cite{Rath2013_PRC88-044617}              \\
              &              &         &         & $0.74$         & $^{159}$Tb 
\cite{Broda1975_NPA248-356}              \\
              &              &         &         & $0.70$         & $^{165}$Ho 
\cite{Tripathi2002_PRL88-172701,Tripathi2005_PRC72-017601}          \\
              &              &         &         & $0.85\pm 0.04$ & $^{197}$Au 
\cite{Palshetkar2014_PRC89-024607}      \\
              &              &         &         &                & $^{198}$Pt 
\cite{Shrivastava2013_PLB718-931}       \\
              &              &         &  &$0.74^{+0.03}_{-0.02}$ &$^{209}$Bi 
\cite{Dasgupta2004_PRC70-024606}       \\
 $^{9}$Be     & $1.573$      & $0.68$  & $0.612$ & $0.80\pm 0.04$ & $^{89}$Y 
\cite{Palshetkar2010_PRC82-044608}    \\
              &              &         &         & $0.90$         & $^{144}$Sm 
\cite{Gomes2006_PRC73-064606,Gomes2009_NPA828-233}         \\
              &              &         &         & $0.72$         & $^{124}$Sn
\cite{Parkar2010_PRC82-054601}        \\
              &              &         &         & $0.60$         & $^{186}$W 
\cite{Fang2013_PRC87-024604}          \\
              &              &         &  &$0.70^{+0.08}_{-0.07}$ & $^{208}$Pb 
\cite{Dasgupta2004_PRC70-024606}      \\
              &              &         &         & $0.68$         &  $^{209}$Bi 
\cite{Signorini1998_EPJA2-227,Dasgupta2010_PRC81-024608}          \\  
 $^{10}$B     & $4.461$      & $0.80$  & $0.799$ & $0.86$         & $^{159}$Tb 
\cite{Mukherjee2006_PLB636-91}      \\
              &              &         &         & $0.85$         & $^{209}$Bi 
\cite{Gasques2009_PRC79-034605}     \\
 $^{11}$B     & $8.665$      & $0.91$  & $0.916$ &                & $^{159}$Tb 
\cite{Mukherjee2006_PLB636-91}      \\
              &              &         &         & $0.93$         & $^{209}$Bi 
\cite{Gasques2009_PRC79-034605}     \\
 $^{12}$C     & $7.367$      & $0.88$  & $0.890$ &                & $^{89}$Y 
\cite{Palshetkar2010_PRC82-044608}  \\
              &              &         &         &                & $^{152}$Sm 
\cite{Broda1975_NPA248-356}         \\
              &              &         &         &                & $^{159}$Tb
\cite{Yadav2012_PRC85-034614}       \\
              &              &         &         &                & $^{181}$Ta
\cite{Babu2003_JPG29-1011}          \\
              &              &         &         & $0.88$         & $^{208}$Pb 
\cite{Kalita2011_JPG38-095104}       \\
 $^{13}$C     & $10.648$     & $0.94$  & $0.943$ &                & $^{159}$Tb
\cite{Yadav2012_PRC86-014603}              \\
              &              &         &         &                & $^{181}$Ta
\cite{Babu2003_JPG29-1011}          \\
              &              &         &         & $0.97$         & $^{207}$Pb 
\cite{Kalita2011_JPG38-095104}       \\
 $^{16}$O     & $7.162$      & $0.87$  & $0.885$ &                & $^{103}$Rh
\cite{Gupta2008_NPA811-77}              \\
              &              &         &         &                & $^{148}$Nd
\cite{Broda1975_NPA248-356}          \\
              &              &         &         &                & $^{150}$Nd
\cite{Broda1975_NPA248-356}          \\
              &              &         &         &                & $^{159}$Tb
\cite{Singh2008_PRC77-014607}       \\
              &              &         &         &                & $^{169}$Tm 
\cite{Singh2008_PRC77-014607}       \\
\end{tabular}\label{tab:SF}
\end{ruledtabular}
 
\end{table}

\subsection{Complete fusion functions for reactions involving weakly and 
tightly bound projectiles}
\label{subsec:com}

The CF functions for the weakly bound projectile ${}^{6}$Li on different target 
nuclei as a function of $x$ are illustrated in Fig. \ref{fig:Li6}. The most 
favorable breakup channel for ${}^{6}$Li is ${}^{6}{\rm 
Li} \rightarrow \alpha + d $ owing to the lowest separation 
energy of $1.474$ MeV. The solid line represents the UFF, i.e., $F_0(x)$ given 
in Eq. (\ref{eq:3}). On the one hand, it can be seen from Fig. \ref{fig:Li6} 
that all of CF functions are below the UFF at energies above the 
Coulomb barrier and one can conclude that the CF cross sections are suppressed 
at 
energies above the Coulomb barrier compared with the UFF. On the other hand, it 
seems that the suppression is almost independent of the target, at least for 
masses larger than $90$.  We introduce a 
suppression factor $F_\mathrm{B.U.}$ as,
\begin{equation}\label{eq:4}
F_\mathrm{B.U.} = \frac{F(x)}{F_0(x)}.
\end{equation}
$F_\mathrm{B.U.}$ is obtained 
by fitting the experimental fusion functions $F(x)$ with $x>0$. In Fig. 
\ref{fig:Li6}, the dotted line represents the UFF multiplied by the 
$F_\mathrm{B.U.}$ of $0.6$. The $F_\mathrm{B.U.}$ of $0.6$ is 
consistent with the results taken from the literature which are listed in the 
fifth column of Table \ref{tab:SF}. These results were obtained by comparing 
the CF data with the coupled channel calculations. One can find that all the 
CF functions are very close to the dotted line, which implies that the 
suppression 
factors for the ${}^{6}$Li projectile  with different targets are almost the 
same. Consequently, the suppression factors for ${}^{6}$Li induced reactions 
are independent of the charge of the target nuclei.

It is interesting to observe that although it has already been shown  
\cite{Otomar2013_PRC87-014615,Hussein2013_PRC88-047601} that the nuclear and 
Coulomb breakups of ${}^{6}$Li increase with the target mass and charge, 
respectively, the effect of the breakup on the complete fusion is roughly 
target 
independent. The reason seems to be related with  the predominance of delayed 
breakups over prompt breakup, where the former is the sequential breakup which 
occurs in two steps. The first step is the excitation of the projectile to a 
long-lived resonance above the breakup threshold. Then, the resonance decays 
into the breakup channel, when the projectile is already in an outgoing 
trajectory leaving the target region. Only prompt breakup, which occurs in a 
time scale of ${10}^{-22}$ s, may affect fusion, since the resonance life-time 
is much longer than the collision time. Actually, Santra {\it et al.}  
\cite{Santra2009_PLB677-139} performed exclusive measurements for the breakup 
of 
${}^{6}$Li in collisions with ${}^{209}$Bi and they found that the sequential 
breakup via the  ${}^{6}$Li $3^{+}$ resonant state at $2.186$ MeV, with 
$T_{1/2}=2.7\times{10}^{-20}$ sec., predominates in the $\alpha$ + $d$ 
fragmentation.

\begin{figure}
\centering{ \includegraphics[width=0.85\columnwidth]{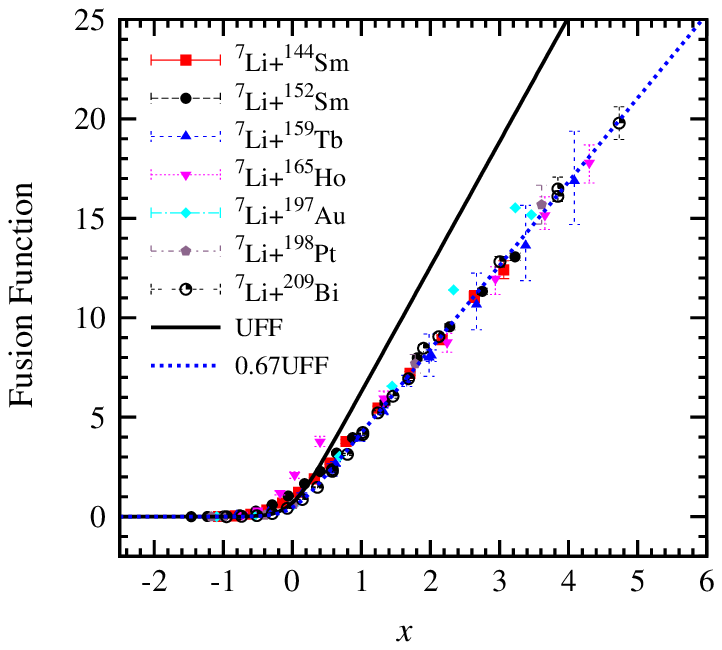}}
\caption{(Color online) The complete fusion function $F(x)$ for the weakly 
bound projectile ${}^{7}$Li on different target nuclei as a function of
$x$. The solid line represents the UFF [Eq. (\ref{eq:3})] and the dotted line is
the UFF multiplied by the suppression factor $F_\mathrm{B.U.}= 0.67$ [cf. Eq. 
(\ref{eq:4})]. The experimental values are taken from 
Refs.~\cite{Rath2013_PRC88-044617}       (${}^{144,152}$Sm),
\cite{Broda1975_NPA248-356}        (${}^{159}$Tb), 
\cite{Tripathi2002_PRL88-172701}   (${}^{165}$Ho),
\cite{Palshetkar2014_PRC89-024607} (${}^{197}$Au),
\cite{Shrivastava2013_PLB718-931}  (${}^{198}$Pt), and
\cite{Dasgupta2004_PRC70-024606}   (${}^{209}$Bi).}\label{fig:Li7}  
\end{figure}

The lowest breakup threshold of the weakly bound projectile ${}^{7}$Li is 
$2.467$ MeV for the breakup channel of ${}^{7}{\rm Li} \rightarrow \alpha + t$. 
The CF functions for projectile ${}^{7}$Li on
different target nuclei as a function of $x$ are shown in Fig.~\ref{fig:Li7}. 
The solid line represents the UFF. It can be seen from Fig.
\ref{fig:Li7} that the CF functions are suppressed owing to the presence of 
the breakup process, which is similar to the results of ${}^{6}$Li. The dotted 
line is the UFF multiplied by the $F_\mathrm{B.U.}$ of $0.67$. The  
$F_\mathrm{B.U.}$ is a little smaller than the results from the literature 
which are shown in the fifth column of the Table \ref{tab:SF}. Those results 
were obtained by comparing the CF cross sections with coupled channel 
calculations, except for the reaction for ${}^{7}$Li on ${}^{165}$Ho  which was 
compared with the predictions of a SBPM 
\cite{Tripathi2002_PRL88-172701,Tripathi2005_PRC72-017601}. From 
Fig.~\ref{fig:Li7}, one can find that all the CF functions 
coincide with the dotted line. So, the suppression factors for the projectile 
${}^{7}$Li induced reactions are also independent of the target charge.

A similar explanation used for ${}^{6}$Li for this behavior can be made for 
${}^{7}$Li. Shrivastava {\it et al.} \cite{Shrivastava2006_PLB633-463} made 
exclusive measurements of ${}^{7}$Li breakup, in collisions with ${}^{65}$Cu. 
They found that the yield of coincident alpha-deuteron was much larger than 
that for coincidences between alpha and triton, and the analysis of angular 
distributions provided clear evidence that the alpha-deuteron events arise from 
a two-step process: direct one-neutron stripping, leaving the projectile in the 
$3^{+}$ resonance of ${}^{6}$Li, followed by its decay into an alpha particle 
plus a deuteron. They concluded that the cross section for the two-step breakup 
process was much larger than that for the direct breakup.

\begin{figure}
\centering{ \includegraphics[width=0.85\columnwidth]{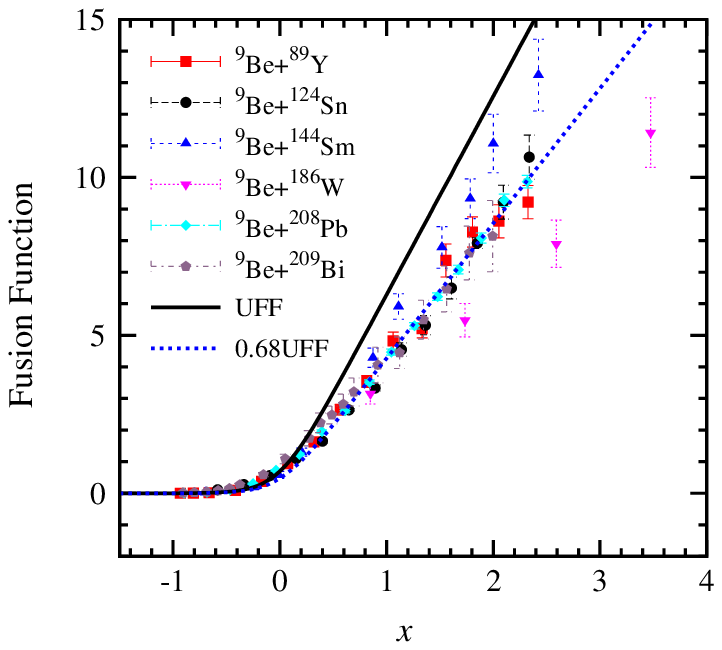}}
\caption{(Color online) The complete fusion function $F(x)$
for the weakly bound projectile ${}^{9}$Be on
different target nuclei as a function of $x$. The solid line
represents the UFF [Eq. (\ref{eq:3})] and dotted line is the 
UFF multiplied by the suppression factor $F_\mathrm{B.U.}= 0.68$ [cf. Eq. 
(\ref{eq:4})]. The experimental values are taken from 
Refs.~\cite{Palshetkar2010_PRC82-044608}                       (${}^{89}$Y),
\cite{Gomes2006_PRC73-064606,Gomes2009_NPA828-233}       (${}^{144}$Sm),
\cite{Parkar2010_PRC82-054601}                           (${}^{124}$Sn),
\cite{Fang2013_PRC87-024604}                             (${}^{186}$W),
\cite{Dasgupta2004_PRC70-024606}                         (${}^{208}$Pb), and
\cite{Signorini1998_EPJA2-227,Dasgupta2010_PRC81-024608} (${}^{209}$Bi). 
}\label{fig:Be9}
\end{figure}

The CF functions for the weakly bound projectile ${}^{9}$Be on 
targets ${}^{89}$Y \cite{Palshetkar2010_PRC82-044608}, ${}^{144}$Sm 
\cite{Gomes2006_PRC73-064606,Gomes2009_NPA828-233}, ${}^{124}$Sn 
\cite{Parkar2010_PRC82-054601}, ${}^{186}$W \cite{Fang2013_PRC87-024604},
${}^{208}$Pb \cite{Dasgupta2004_PRC70-024606}, and ${}^{209}$Bi
\cite{Signorini1998_EPJA2-227,Dasgupta2010_PRC81-024608} are plotted in Fig. 
\ref{fig:Be9}. The lowest breakup threshold of ${}^{9}$Be is $1.573$ MeV for its
breakup into $n + \alpha + \alpha$. In Fig. 
\ref{fig:Be9}, the solid line 
represents the UFF, and the dotted line denotes the UFF scaled by the 
$F_\mathrm{B.U.}$ of $0.68$. The data of ${}^{9}$Be on 
${}^{144}$Sm and ${}^{186}$W are not included in the fitting. One can see 
that most of CF functions are very close to the dotted line, which is 
consistent with the results of ${}^{6,7}$Li. The present suppression factor
$F_\mathrm{B.U.}$ is consistent with the factors obtained by comparing the CF 
data with the coupled channel calculations 
\cite{Palshetkar2010_PRC82-044608, 
Parkar2010_PRC82-054601, Dasgupta2004_PRC70-024606} and a SMPB 
\cite{Dasgupta2010_PRC81-024608}. It seems that the 
suppression factor should be smaller for ${}^{9}$Be on ${}^{144}$Sm and 
larger for ${}^{9}$Be on ${}^{186}$W.  
The systematic behavior of CF suppression of the weakly bound projectile 
${}^{9}$Be incident ${}^{144}$Sm, ${}^{168}$Er, 
${}^{186}$W, ${}^{196}$Pt, ${}^{208}$Pb, and ${}^{209}$Bi was also explored 
based on a three-body classical trajectory model with stochastic breakup in
Ref.~\cite{Rafiei2010_PRC81-024601}. The authors suggested that the 
discrepancy between ${}^{144}$Sm and other targets may be attributed to the 
fact that the measured ICF cross section was only a lower limit and 
concluded that the suppression factor is nearly independent of the target 
charge. 
In Ref.~\cite{Gomes2011_PRC84-014615}, the systematic behavior of CF 
suppression 
of the weakly bound projectile ${}^{9}$Be incident ${}^{89}$Y, ${}^{124}$Sn, 
${}^{144}$Sm, and ${}^{208}$Pb was also investigated by comparing with the 
UFF, it was concluded that the systematic behavior for the CF 
suppression as a function of the target charge is not clear, which may be 
explained by different effects of transfer channels, especially one-neutron 
stripping, on the CF or TF.

Again, for ${}^{9}$Be, Hinde {\it et al.} \cite{Hinde2002_PRL89-272701} 
performed coincidence experiments to determine time scales in the breakup of 
${}^{9}$Be, in collisions with a ${}^{208}$Pb target at sub-barrier energies. 
They were able to disentangle prompt ${}^{9}$Be breakup from the delayed 
breakup 
of ${}^{8}$Be, triggered by a one-neutron stripping process. In the latter 
case, 
the transfer reaction produces the unstable ${}^{8}$Be nucleus, which has the 
half-life $T_{1/2}= {10}^{-16}$ s, several orders of magnitude longer than the 
collision time, and so a process of this kind cannot influence fusion.

\begin{figure}
\centering{ \includegraphics[width=0.85\columnwidth]{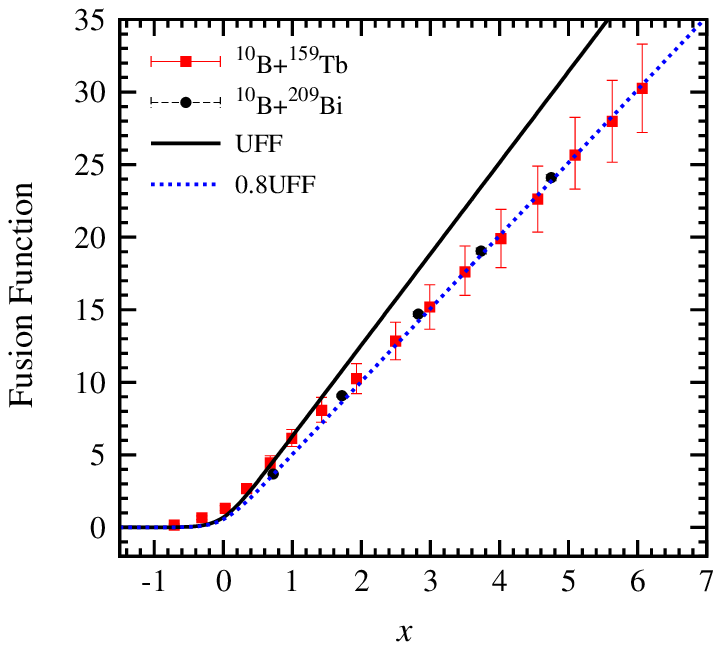}}
\caption{(Color online) The complete fusion function $F(x)$ for the weakly bound
projectile ${}^{10}$B on different target nuclei as a function of $x$. The
solid line represents the UFF [Eq. (\ref{eq:3})] and the dotted line is 
the UFF multiplied by the suppression factor $F_\mathrm{B.U.}= 0.8$ [cf. Eq. 
(\ref{eq:4})]. The experimental data are from 
Refs.~\cite{Mukherjee2006_PLB636-91}  (${}^{159}$Tb) and 
\cite{Gasques2009_PRC79-034605} (${}^{209}$Bi).}\label{fig:B10}
\end{figure}

Figure~\ref{fig:B10} shows the CF functions for the weakly bound projectile
${}^{10}$B on ${}^{159}$Tb and ${}^{209}$Bi targets. The data 
for ${}^{10}$B with ${}^{159}$Tb and ${}^{209}$Bi are taken 
from Refs. \cite{Mukherjee2006_PLB636-91} and \cite{Gasques2009_PRC79-034605},
respectively. The most favorable breakup channel for ${}^{10}$B is 
${}^{10}{\rm B} \rightarrow {}^{6}{\rm Li} + \alpha$, of which 
the breakup threshold is $4.461$ MeV. The CF functions for ${}^{10}$B 
on target ${}^{159}$Tb coincide with those for target ${}^{209}$Bi. 
The CF functions lie below the UFF, as expected, 
which can be assigned to the breakup effects on the fusion process. The 
suppression factor is obtained as $0.8$ by making a fit.  This 
$F_\mathrm{B.U.}$ is a little smaller than that of $0.85$ which 
is obtained by comparing the CF cross sections with 
the predictions of a SBPM \cite{Gasques2009_PRC79-034605}. One can 
find that $F_\mathrm{B.U.}$ for ${}^{10}$B is larger than those for 
${}^{6,7}$Li and ${}^{9}$Be, because ${}^{10}$B has a 
larger breakup threshold.

\begin{figure}
\centering{\includegraphics[width=0.85\columnwidth]{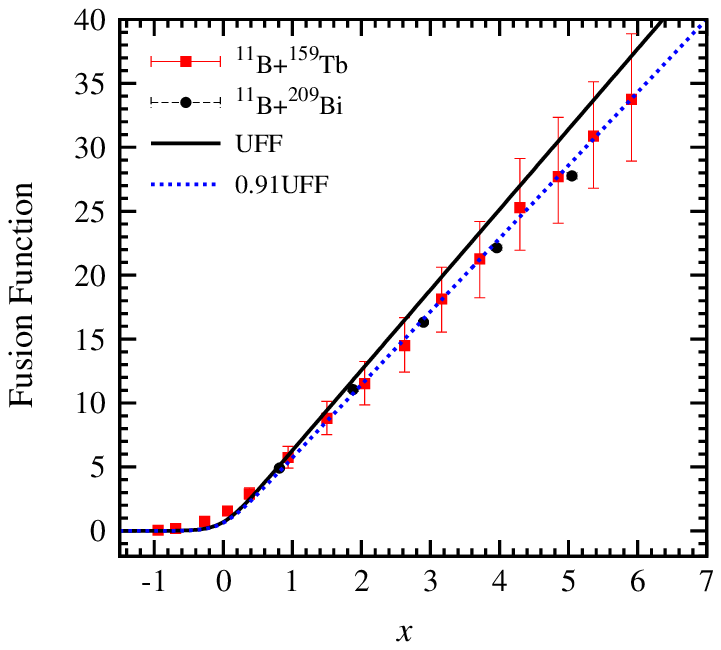}}
\caption{(Color online) The complete fusion function $F(x)$ for the tightly 
bound projectile ${}^{11}$B on different target nuclei as a function of $x$. The
solid line represents the UFF [Eq. (\ref{eq:3})] and the dotted line is 
the UFF multiplied by the suppression factor $F_\mathrm{B.U.}= 0.91$ [cf. Eq. 
(\ref{eq:4})]. The 
experimental values are from 
Refs.~\cite{Mukherjee2006_PLB636-91} (${}^{159}$Tb) and 
\cite{Gasques2009_PRC79-034605} (${}^{209}$Bi).}\label{fig:B11}
\end{figure}

Next we further investigate the effects of breakup coupling on CF reactions
with tightly bound nucleus as a projectile.
The CF functions for projectile ${}^{11}$B on ${}^{159}$Tb and 
${}^{209}$Bi targets are shown in Fig. \ref{fig:B11}. The data are taken 
from Refs. \cite{Mukherjee2006_PLB636-91} for ${}^{159}$Tb and 
\cite{Gasques2009_PRC79-034605} for ${}^{209}$Bi,
respectively. For the nuclide ${}^{11}$B, the breakup threshold energy is 
$8.665$ MeV for the most favorable breakup channel of ${}^{11}{\rm B} 
\rightarrow {}^{7}{\rm Li} + \alpha$. From the comparison between the 
CF functions and the UFF, it is found that the CF functions coincide with the 
UFF scaled by the $F_\mathrm{B.U.}$ of $0.91$ which is displayed by the dotted 
line. Comparing with the results for its neighboring nuclide ${}^{10}$B, we 
find that the suppression factor is larger, as well as the breakup threshold, 
which is similar to the cases of ${}^{6,7}$Li. 
  
\begin{figure}
\centering{ \includegraphics[width=0.85\columnwidth]{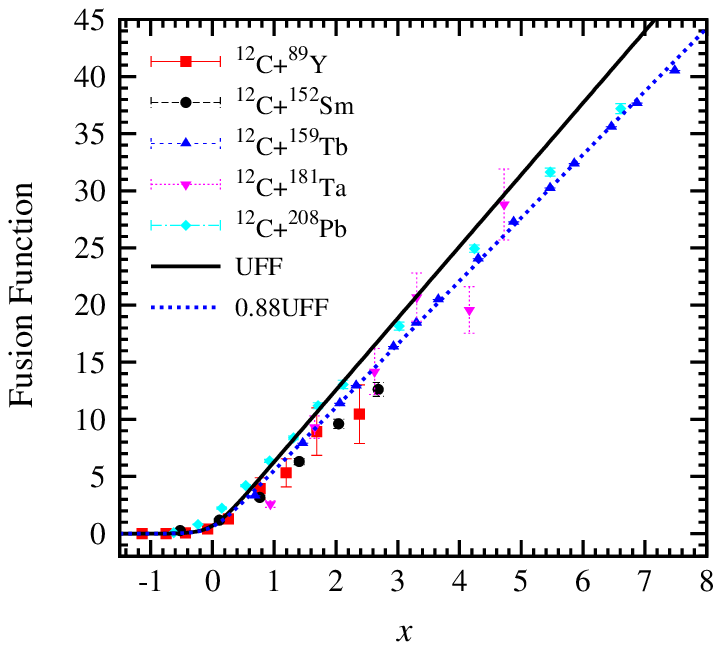}}
\caption{(Color online) The complete fusion function $F(x)$ for tightly bound 
projectile ${}^{12}$C on different target nuclei as a function of $x$. The 
solid 
line represents the UFF [Eq. (\ref{eq:3})] and the dotted line is the UFF 
multiplied by the suppression factor $F_\mathrm{B.U.}= 0.88$ [cf. Eq. 
(\ref{eq:4})]. The experimental values are taken from 
Refs.~\cite{Palshetkar2010_PRC82-044608}   (${}^{89}$Y),
\cite{Broda1975_NPA248-356}          (${}^{152}$Sm), 
\cite{Yadav2012_PRC85-034614}        (${}^{159}$Tb),
\cite{Babu2003_JPG29-1011}           (${}^{181}$Ta), and
\cite{Kalita2011_JPG38-095104}       (${}^{208}$Pb).}\label{fig:C12}
\end{figure}
 
\begin{figure}
\centering{ \includegraphics[width=0.85\columnwidth]{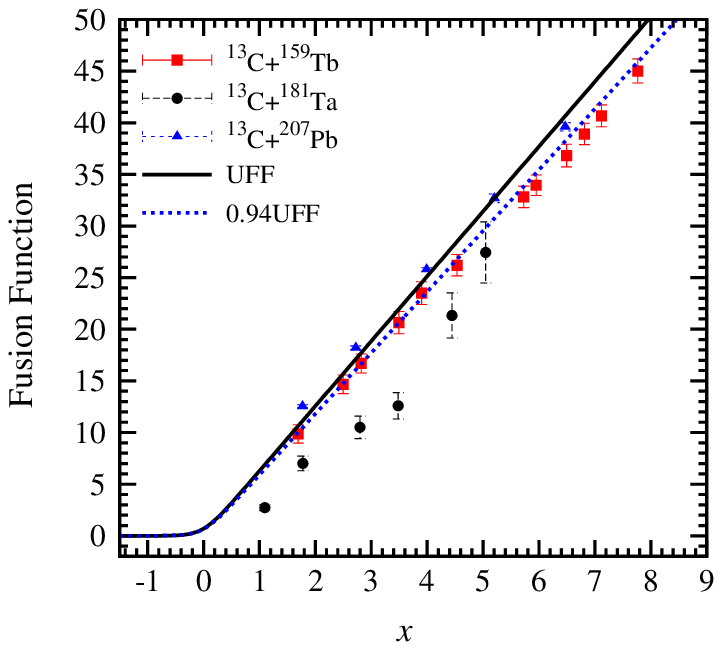}}
\caption{(Color online) The complete fusion function $F(x)$ for tightly bound 
projectile ${}^{13}$C on different target nuclei as a function of $x$. The 
solid 
line represents the UFF [Eq. (\ref{eq:3})] and the dotted line is the 
UFF multiplied by the suppression factor $F_\mathrm{B.U.}= 0.94$ [cf. Eq. 
(\ref{eq:4})]. The experimental values are taken from 
Refs.~\cite{Yadav2012_PRC86-014603}        (${}^{159}$Tb),
\cite{Babu2003_JPG29-1011}           (${}^{181}$Ta), and
\cite{Kalita2011_JPG38-095104}       (${}^{207}$Pb).}\label{fig:C13}
\end{figure}

The CF functions of reactions with ${}^{12}$C as a projectile are illustrated 
in Fig.~\ref{fig:C12}. The lowest energy breakup channel is
${}^{12}$C $\rightarrow {}^{8}{\rm Be} + \alpha$ with a
threshold energy $7.367$ MeV. It can be seen that the CF functions 
%, except for ${}^{12}$C incident ${}^{181}$Ta, 
are close to the UFF
multiplied by the $F_\mathrm{B.U.}$ of $0.88$. It is identical to 
the factor obtained for ${}^{12}$C on ${}^{208}$Pb by comparing with the 
predictions of a SBPM \cite{Kalita2011_JPG38-095104}. The suppression factor 
is larger than that of ${}^{10}$B and smaller than that of ${}^{11}$B, which is 
related to that the breakup threshold energy of ${}^{12}$C is larger than that 
of 
${}^{10}$B and smaller than that of ${}^{11}$B. 

The CF data for the tightly bound projectile ${}^{13}$C are also
studied and the CF functions are shown in Fig.~\ref{fig:C13}.
For ${}^{13}$C, the most favorable breakup channel is ${}^{13}{\rm C} 
\rightarrow {}^{9}{\rm Be} + \alpha$, with a threshold energy 
of $10.648$ MeV. The $F(x)$ for ${}^{13}{\rm C}$ with ${}^{181}{\rm 
Ta}$ is far below the UFF and not used in the 
fitting for the suppression factor. As expected, the suppression factor 0.94 is 
larger than that of 
${}^{12}$C and ${}^{11}$B because of its larger threshold energy than those of 
${}^{12}$C and ${}^{11}$B. 
%One can find that the $F_\mathrm{B.U.}$ of $0.94$ is 
%larger than those of ${}^{12}$C and ${}^{11}$B which is consistent with the 
%expected result.

\begin{figure} 
\centering{ \includegraphics[width=0.85\columnwidth]{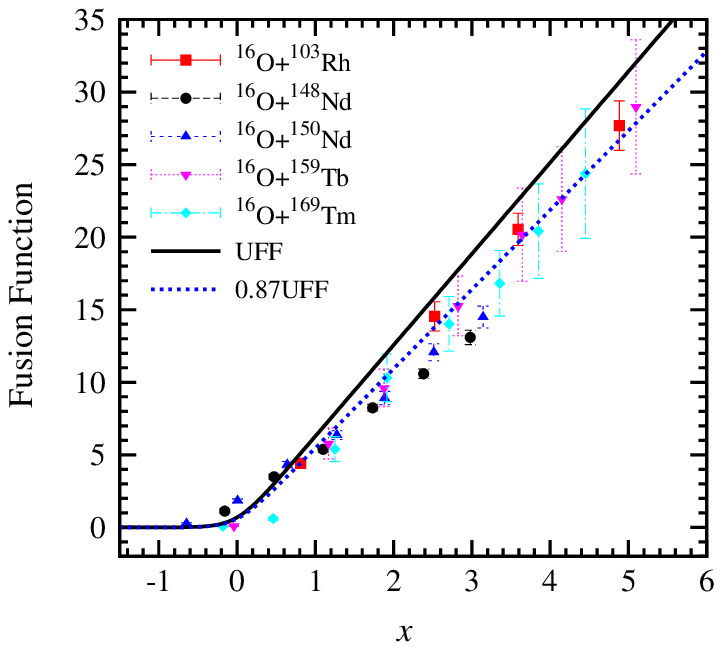}}
\caption{(Color online) The complete fusion function $F(x)$ for tightly bound 
projectile ${}^{16}$O on different target nuclei as a function of $x$. The 
solid 
line represents the UFF [Eq. (\ref{eq:3})] and the dotted line is the UFF 
multiplied by the suppression factor $F_\mathrm{B.U.}= 0.87$ [cf. Eq. 
(\ref{eq:4})]. The experimental values are taken from 
Refs.~\cite{Gupta2008_NPA811-77}    (${}^{103}$Rh),
\cite{Broda1975_NPA248-356}   (${}^{148,150}$Nd), and
\cite{Singh2008_PRC77-014607} (${}^{159}$Tb, ${}^{169}$Tm).}\label{fig:O16}  
\end{figure}

Figure~\ref{fig:O16} shows the CF functions for reactions with the tightly bound
projectile ${}^{16}$O. The threshold energy of ${}^{16}$O is
$7.162$ MeV for the lowest energy breakup channel, ${}^{16}{\rm O} \rightarrow 
{}^{12}{\rm 
C} + \alpha$. The suppression factor is smaller 
than that of ${}^{12}$C because of a lower breakup threshold compared to 
${}^{12}$C. From Fig.~\ref{fig:O16}, it can be seen that the CF 
functions of ${}^{16}{\rm O}$ on ${}^{103}$Rh, ${}^{159}$Tb, and ${}^{169}$Tm 
are very close to the dotted line which is the UFF multiplied by the 
$F_\mathrm{B.U.}$ of $0.87$. The suppression effect is also independent of 
the target charge, which is consistent with the results of ${}^{6,7}$Li, 
${}^{10,11}$B, ${}^{9}$Be, and ${}^{12}$C. The results of CF functions of 
${}^{16}$O + ${}^{148,150}$Nd, which are not included in the 
fitting, lie far below the dotted line. 
Actually the TF functions also lie below the UFF. 
We expect new experimental investigations of these reactions.
 
\begin{figure}
\centering{ \includegraphics[width=0.95\columnwidth]{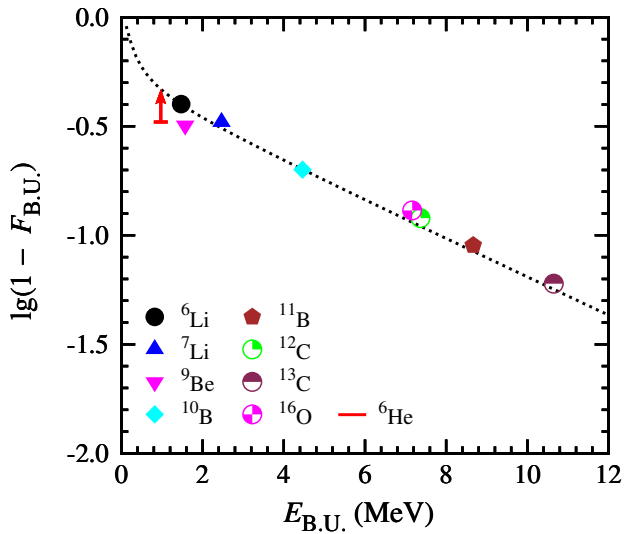}}
\caption{\label{modification:Fig9} (Color online) The suppression factors for 
projectiles ${}^{6,7}$Li, 
${}^{9}$Be,
${}^{10,11}$B, ${}^{12,13}$C, and ${}^{16}$O as a function of the lowest 
projectile breakup threshold ($E_{\rm B.U.}$) for reactions on different 
targets. The solid line represents the suppression factor of total
fusion for reactions with ${}^{6}$He as a projectile. 
The dotted line denotes empirical formula (\ref{eq:E_BU}). 
}\label{fig:E_BU}
\end{figure}
 
\subsection{Systematics of suppression factors}
\label{subsec:sys}

Based on the above analysis and discussions, one can conclude that the 
CF suppression for the reactions induced by the same 
nuclide is independent of the target charge. 
The suppression factors and the 
breakup threshold energies for ${}^{6,7}$Li, ${}^{9}$Be,
${}^{10,11}$B, ${}^{12,13}$C, and ${}^{16}$O projectiles are listed in 
Table~\ref{tab:SF}. One can find that the suppression factor is 
sensitive to the breakup threshold energy of the projectile. Therefore, it is 
necessary to explore the relation between the suppression factor and the 
breakup 
threshold. This has been done by investigating the reactions with lead and 
bismuth targets~\cite{Gasques2009_PRC79-034605,Kalita2011_JPG38-095104}. 
We will study this question with more reaction systems.

The suppression factors for projectiles 
${}^{6,7}$Li, ${}^{9}$Be, ${}^{10,11}$B, ${}^{12,13}$C, and ${}^{16}$O are 
shown as a function of the lowest projectile breakup threshold in 
Fig.~\ref{fig:E_BU}. One can see that ${}^{6}$Li induced reactions have the 
strongest suppression ($F_{\rm B.U.}=0.6$), and ${}^{13}$C has the largest 
suppression factor of $0.94$. This is because of ${}^{6}$Li having the lowest 
breakup threshold of $1.474$ MeV and ${}^{13}$C having the highest breakup 
threshold of $10.648$ MeV. 
\label{modification:FbuEbu} 
Figure~\ref{fig:E_BU} shows that there exists an 
exponential relation between the suppression factor and the breakup threshold 
energy, at least for 1.474 MeV $<E_{\rm B.U.}<$ 10.648 MeV.
Furthermore, if the breakup threshold energy becomes even larger, 
breakup effects would play an even smaller role and $F_{\rm B.U.}$ should be close to $1$.
On the contrary, when the breakup 
threshold energy is negligible, the interaction (specially the
long range Coulomb force) will break the projectile at very large distances.
So no complete fusion will occur, i.e., $F_{\rm B.U.}\approx0$. 
An analytical formula that fulfills these physical limits is
\begin{equation}\label{eq:5}
 \lg(1-F_{\rm B.U.}) =  -a\exp(-b/E_{\rm B.U.})-cE_{\rm B.U.},
\end{equation}
where $a$, $b$, and $c$ are parameters to be determined. 
By fitting the suppression factors given in 
Table~\ref{tab:SF} and shown in Fig.~\ref{fig:E_BU}, 
we get the values for the three parameters, 
$a=0.33$, $b=0.29$ MeV and $c=0.087~\text{MeV}^{-1}$. 
That is, this analytical formula reads
\begin{equation}\label{eq:E_BU}
 \lg(1-F_{\rm B.U.}) = -0.33\exp(-0.29/E_{\rm B.U.})-0.087E_{\rm B.U.},
\end{equation}
or equivalently,
\begin{equation}\label{eq:6}
 \ln(1-F_{\rm B.U.}) = -0.76\exp(-0.29/E_{\rm B.U.})-0.2E_{\rm B.U.},
\end{equation}
where $E_{\rm B.U.}$ is in the unit of MeV.
The suppression factors obtained by these empirical formulas are also listed in 
Table~\ref{tab:SF} and shown in Fig.~\ref{fig:E_BU} as a dotted line.
From Fig.~\ref{fig:E_BU}, one can find that the  
$F_{\rm B.U.}$ for ${}^{9}$Be is a little larger than that suggested from the 
empirical formula. 
This analytical relation suggests that the influence of the breakup channel 
on the complete fusion is a threshold effect.
The physics behind it is still unclear. 
Further experimental and 
theoretical studies are expected.

These conclusions may look, at a first sight, contradictory with recent 
experimental evidences that the sequential breakup of the weakly bound nuclei 
${}^{6,7}$Li and ${}^{9}$Be, following neutron and proton transfer 
\cite{Shrivastava2006_PLB633-463,Rafiei2010_PRC81-024601,
Luong2011_PLB695-105,Luong2013_PRC88-034609} predominates over the direct 
breakup, 
at least at sub-barrier energies. If this is so, one might not expect such 
clear dependence of the complete fusion with the direct breakup threshold 
energy 
as shown in Eqs.~(\ref{eq:5}-\ref{eq:6}). However, as we have pointed out 
previously, this sequential breakup is of the delayed type and cannot affect 
fusion. Therefore, the effect of breakup on fusion may indeed depend on the 
breakup threshold energy.

Finally we focus on a typical neutron halo nucleus, ${}^{6}$He.
The lowest energy breakup channel of ${}^{6}$He is $\alpha + 2n$
with a threshold energy 0.972 MeV.
Experimentally, only total fusion cross sections have been measured 
\cite{Kolata1998_PRL81-4580,Kolata1998_PRC57-6R}.
It is interesting to note that, with the UFF as a standard reference,
the TF of reaction systems with ${}^{6}$He as a projectile is also 
suppressed by the breakup and the TF suppression factor
$F^\mathrm{TF}_\mathrm{B.U.}$ is 0.67 \cite{Canto2014_in-prep} 
(cf. Ref.~\cite{Canto2009_NPA821-51} where this suppression factor was 0.7).
This TF suppression factor is also shown in Fig.~\ref{fig:E_BU}. 
One can find that $\lg(1-F^\mathrm{TF}_\mathrm{B.U.})$ of ${}^{6}$He 
is below the prediction of the empirical formula (\ref{eq:E_BU}). 
Since CF cross sections must be smaller than TF cross sections, 
the CF suppression factor for ${}^{6}$He should be smaller than 
the TF suppression factor. 
That is, 0.67 can only be treated as an upper limit of the CF suppression
factor for ${}^{6}$He and $\lg(1-F_\mathrm{B.U.})$ should 
be closer to that from the empirical formula (\ref{eq:E_BU}). 
More efforts should be devoted to measuring the CF cross sections
reaction systems with ${}^{6}$He as a projectile.

\section{Summary}
\label{sec:summary}

In order to investigate the influence of breakup on the complete fusion (CF) at 
energies above the Coulomb barrier, we adopt the double folding and 
parameter-free 
S\~ao Paulo potential to get the barrier parameters of the reactions induced by 
the weakly or 
tightly bound projectiles. The barrier parameters are used to extract the 
dimensionless fusion functions $F(x)$ from the CF cross sections for 
${}^{6,7}$Li, ${}^{9}$Be, ${}^{10,11}$B, ${}^{12,13}$C, and ${}^{16}$O induced 
reactions. Then the fusion functions $F(x)$ are compared with the 
universal fusion function (UFF). 
From the fact that the fusion function $F(x)$ is always below the UFF, 
we conclude that the CF cross sections are suppressed owning to the 
prompt breakup of projectiles. 
The CF suppression for the reactions induced by the same projectile is 
independent of target charge. 
The suppression factors for different projectiles are mainly 
determined by the lowest breakup thresholds.
Based on the systematics obtained in this work, 
we propose an analytical formula which describes well the relation
between the CF suppression factor and the breakup threshold energy. 
%The physics behind this exponential 
%relation is still unclear and further experimental and theoretical 
%investigations are expected.

\acknowledgements
Helpful discussions with Mahananda Dasgupta, Alexis Diaz-Torres, R. V. Jolos, 
Kai Wen, Huan-Qiao Zhang, Zhen-Hua Zhang, and Jie Zhao are gratefully 
acknowledged. 
We thank the referee for suggestions on discussions about 
the Wong's formula and the asymptotic behaviors of the analytical
expression for the suppression factor.
This work also benefited from discussions held at CUSTIPEN 
(China-U.S. Theory Institute for Physics with Exotic Nuclei).
P.R.S.G. acknowledges the 
partial financial support from CNPq, FAPERJ, and the PRONEX.
This work has been partly supported by 
the National Key Basic Research Program of China (Grant No. 2013CB834400), 
the National Natural Science Foundation of China (Grants 
No. 11121403, 
No. 11175252, 
No. 11120101005, 
No. 11211120152, and 
No. 11275248),
and
the Knowledge Innovation Project of the Chinese Academy of Sciences (Grant No. 
KJCX2-EW-N01).
The computational results presented in this work have been obtained on 
the High-performance Computing Cluster of SKLTP/ITP-CAS and 
the ScGrid of the Supercomputing Center, Computer Network Information Center of 
the Chinese Academy of Sciences.
 
%\bibliography{/home/wb/Dropbox/Mybib.bib}
%\bibliography{../../../information/refs/JabRef/sgzhou}

%merlin.mbs apsrev4-1.bst 2010-07-25 4.21a (PWD, AO, DPC) hacked
%Control: key (0)
%Control: author (8) initials jnrlst
%Control: editor formatted (1) identically to author
%Control: production of article title (-1) disabled
%Control: page (0) single
%Control: year (1) truncated
%Control: production of eprint (0) enabled
%

\end{document}